\documentclass{aa}  
\usepackage{myStyle}
\usepackage{graphicx}
\usepackage{txfonts}
\begin{document}
   \title{Precessing planetary magnetospheres in SiO stars~?}
   \subtitle{First detection of quasi-periodic polarization fluctuations
             in R\,Leo and V\,Cam}

   \author{H.~W. Wiesemeyer
          \inst{1}\fnmsep\thanks{On leave to Instituto de Radioastronom\'ia Milim\'etrica, Granada, Spain.}
          \and
          C. Thum\inst{1}
          \and
          A. Baudry \inst{2,3} 
          \and
          F. Herpin \inst{2,3}
          }

   \offprints{H. Wiesemeyer}

   \institute{Institut de Radio Astronomie Millim\'etrique, \\
              300, Rue de la Piscine, F-38406 Saint Martin d'H\`eres, France
         \and
  Universit\'e de Bordeaux, Laboratoire d'Astrophysique de Bordeaux, \\
  F-33000 Bordeaux, France \and CNRS/INSU, UMR 5804, BP 89, F-33270 Floirac, France}
   \date{Received ; accepted }

 
  \abstract
   {The origin of magnetism around asymptotic giant branch (AGB) stars is still uncertain.
    These stars may drive an important dynamo, but if the magnetic energy dissipates
    all in X-rays, the observed X-ray luminosities are too low for a strong, dynamically 
    important global field. Other explanations for the circular polarization of SiO masers
    in the AGB atmospheres may thus be needed.}
    {The interaction of the AGB wind with previously ejected matter and with planets is expected
    to bear complex magneto-hydrodynamic phenomena on a short time scale, such that 
    strong magnetic fields can be maintained locally. Here we provide observational evidence for the
    corresponding magnetic fluctuations.} 
    {We use the circular polarization of the $v=1,J=2-1$ SiO masers as tracer for magnetic activity.
    A correlation polarimeter allows us to simultaneously record all Stokes parameters. An SiO maser
    survey of 80 AGB stars was performed, out of which eight sources with the strongest circular
    polarization were selected for further monitoring.}
    {In two AGB stars, V~Cam and R~Leo, we find evidence of pseudo-periodic fluctuations
    of the fractional circular polarization on a timescale of a few hours, from which we infer magnetic
    fluctuations of $\sim 1$\,G. The phenomenon is rare and, if detected in an SiO star, restricted to
    a narrow range of velocities. It seems to be associated with planetary wake flows suggested by
    VLBI maps.}
    {While scenarios involving magnetic activity in the extended stellar atmosphere have
    problems to explain all observed features, precessing Jovian magnetospheres predict all
    of them without difficulty. For the case of R\,Leo, we constrain the orbit of the planet
    (estimated period 5.2\,years), derive a stellar mass estimate of $0.7\,M_{\sun}$ from it, and discuss
    the impact of planetary magnetism on the survival of planets. Smooth velocity variations of the fluctuating
    circular polarization feature are predicted as the planet moves along its orbit.}

   \keywords{masers -- polarization -- stars: AGB and post-AGB, atmospheres,
             magnetic field, planetary systems -- techniques: polarimetric}

   \maketitle
%

\section{Introduction}
Towards the end of their lifetime, low- to intermediate mass stars undergo a phase in which they burn Helium in a shell on
top of a Carbon-Oxygen core, and Hydrogen in another shell above the Helium shell. During this phase, when the stars
appear in the Hertzsprung-Russell diagram in the upper asymptotic giant branch (AGB), they lose, due to pulsations, at
least half of their mass, which forms a circumstellar envelope (see Herwig, 2005, for a review). The inner part of this
envelope, also called extended atmosphere, is expected to bear complex magneto-hydrodynamic phenomena, due to the
interaction of the wind with the previously ejected matter and with planets (Struck et al., 2004, 2002, Struck-Marcell,
1988). As in the solar system, where space weather changes on timescales of hours (e.g. Prang\'e et al., 2004),
fluctuations of the magnetic field about a mean value can be expected due to this interaction, but the observational
evidence is still lacking. Here we will show that for a narrow range of velocities the circular polarization of SiO
masers, generally accepted as a tracer of the magnetic field in the extended atmosphere of 
AGB stars (Nedoluha \& Watson, 1994, Elitzur, 1996), varies in two AGB stars with a period of a few hours. Previous
multi-epoch observations (Diamond et al., 2003, Pardo et al., 2004, Kang et al., 2006) of SiO masers were not
polarimetric, while the sampling of the polarization variability study of Glenn et al. (2003) was not dense enough to
detect such intraday magnetic fluctuations.

Our knowledge about magnetism within the extended atmosphere of AGB stars relies on circular polarization measurements of
SiO masers at 1.5 to 7 AU distance from a star with a typical radius of 0.7 to 1\,AU. Maser theory implies molecular
excitation in dense pockets of gas and amplification of light in narrow tubes. Assuming that the circular polarization is
caused by the Zeeman effect in the non-paramagnetic SiO molecules in the tubes, the reported fractional circular
polarizations of up to 9\,\% for the $v=1, J=1-0$ transition ($v$ and $J$ are the vibrational and rotational quantum
numbers, respectively) yield line-of-sight averaged magnetic flux densities of up to 100\,G (Barvainis et al., 1987). More
recent observations (Herpin et al., 2006) of the $v=1, J=2-1$ masers (excited in gas layers very close to those where
$v=1, J=1-0$ is excited) also show the presence of circular and linear polarizations in many stars. These observations
result in lower magnetic flux density estimates in the range 1 to 15\,G, again assuming the circular polarization is due to 
the Zeeman effect. In the outer envelope, H$_2$O and OH masers reside at $\sim 100$ to 400\,AU, respectively $\sim 1000$ to 
$\sim 10000$ AU, from the star. The Zeeman effect of H$_2$O masers has been used (Vlemmings et al., 2002) to estimate, by
extrapolation assuming a solar-type field topology, a magnetic flux density at the stellar surface of 100\,G. In summary,
it is now observationally evident that AGB stars maintain a magnetic flux throughout their envelope. Blackman et al. (2001) 
showed that AGB stars can generate a magnetic field via dynamo action at the interface between the rapidly rotating core
and the extended convection zone. The importance of the magnetic field for global envelope dynamics is still a
matter of debate, though (e.g. Soker, 2006). 

Once the SiO maser spots are formed (often distributed along arcs around the star, hereafter referred to as maser shell,
see e.g. Cotton et al., 2008), they are subject to magnetospheric events which are known to be rapidly variable
if we refer to the solar system. Here we provide first evidence for such fluctuations in the atmospheres of two AGB stars
by frequent polarization sampling. The polarization monitoring by Glenn et al. (2003) was only sensitive to slow variations 
and therefore to a long-term readjustment of the magnetic field (on a timescale of several months).

\section{Observations and data analysis}
\subsection{Instrumentation and observations}
The bulk of our observations was done in May 2006 with the IRAM 30m telescope at Pico Veleta,
which is equipped with dual-polarization receivers. To confirm the rapid polarization variability
discussed here, Mira-type star R\,Leo was observed during additional 4\,hours on 2008 August 01.
The receiver pair operating at 3mm wavelength was tuned to the $v=1, J=2-1$ transition of SiO at 86.243 GHz and made
coherent by using the same local oscillator reference. The signals were analyzed with XPOL (Thum et al., 2008), a
correlation spectrometer enhanced with cross-correlation products for the signals from the orthogonally polarized
receivers, at 39.0625\,kHz or 0.136\,\kms channel spacing within a 16.2\,MHz bandwidth. Receiver temperatures varied
between 36\,K and 64\,K, and system temperatures between 70\,K and 235\,K, resulting in an antenna temperature
($T_{\rm A}^*$) noise of $\sigma_{\rm rms} = 36 - 100$\,mK across the spectral baseline. The conversion factor from
$T_{\rm a}^*$ to the flux density scale is 6\,Jy/K. The phase difference between the receivers was measured with a polarizing grid
mounted in front of the cold load (at about $70-80$\,K effective temperature) of the calibration unit. Whenever the
temperature scale was calibrated (every 10 to 20 minutes), the signal from the polarizer, with well known properties, was
also observed and compared to the unpolarized calibration load at ambient temperature. The phase correction was then
applied in order to attribute the measured real and imaginary part of the cross correlation product to the Stokes U
parameter (in the reference frame of the telescope's Nasmyth focus) and to the Stokes V parameter (positive for a right
hand circularly polarized signal, according to IAU convention). Small residual phase errors result in a leakage of the Stokes U signal
into Stokes V, a subtle effect only worth worrying about when Stokes U is strong (which is the case here).
The scaled ($\sim 3\,\%$) copy of the Stokes U signal thus contaminating the measured circular polarization can be
removed, since it is subject to the parallactic rotation of the polarization vector, which easily shows in the data. The
remaining Stokes V signal is intrinsic, the contribution of the telescope and the receiver cabin optics being negligible
($-0.03 \pm 0.12\,\%$ on the optical axis).

77 AGB stars were observed, out of which 62 are Mira stars, 10 are semiregular variable stars, and 2 red supergiants. Eight
objects with the strongest circular polarization were retained for dense monitoring (twice per hour). They are all Mira stars,
in agreement with the finding of Herpin et al. (2006) that this source class tends to have the strongest linear and circular
polarizations, and is therefore well suited for our aims.
\subsection{Data analysis}
The temperature and phase calibration was done with the MIRA raw data reduction software$^1$. The subsequent processing of the
Stokes spectra, described in the following, was done with the CLASS and GREG software\footnote{GILDAS collection, see
http://www.iram.fr/IRAMFR/GILDAS}. Due to the irregular sampling, a determination of the spectral power density
(SPD) of the time series using Fourier transform techniques would yield poor results, and preference was given to the SPD
estimate provided by the Lomb technique (Lomb, 1976, see also Press et al., 1994). This method only uses the measured data
without any prior interpolation to a regular sampling function, and does a least-square fitting to the harmonic contents of 
the time series. A detection of quasi-periodic fluctuations of the fractional circular polarization (hereafter $p_{\rm C}$) 
needs to pass three critical tests. First, we do not expect a significant fluctuation in the Stokes U residuals left after
the subtraction of the stationary (within the timescales considered here) linear polarization feature. Second, the
oscillation of $p_{\rm C}$ was tested against the null hypothesis that the data are not periodic, but random noise with a
Gaussian distribution; this test yields a false-alarm probability. Third, because a restricted number of random samples
cannot be strictly Gaussian, Monte-Carlo simulations, with random samples contemporaneous with the observed features, were
used to confirm the significance of the false-alarm probabilities.
   \begin{table*}
      \caption[]{Observational Results}
      \label{tab:results}
      \centering
      \begin{tabular}{lll}
          \hline\hline
          \noalign{\smallskip} & V~Cam     &  R~Leo \\
          \hline
          Phase of optical lightcurve$^{(1)}$ & 0.08 & 0.10 \\
          Oscillation period of circular polarization$^{(2)}$
                     & $(5.4\pm 0.1)$\,h & $(6.3 \pm 0.3)$\,h \\
          False-alarm probability$^{(3)}$ & 4\,\% & 16\,\%  \\
          Radial velocity of maser spots$^{(4)}$ 
                     & 7.5\,\kms & 4.4\,\kms \\
          Mean circular polarization
                     & $-0.3\,\%$ & $+2.0\,\%$ \\
          Mean magnetic flux density in maser$^{(5)}$
                     & $\sim 100$\,mG & $\sim 1$\,G \\
          Peak-to-peak variation of fractional polarization 
                     & $2.8\,\% = 5.6\,\sigma_{\rm rms}$
                     & $3.7\,\% = 7.4\,\sigma_{\rm rms}$  \\
          Fluctuation of magnetic field$^{(5)}$
                     & $\sim 1$\,G & $\sim 1$\,G \\
            \hline
      \end{tabular}
\begin{list}{}{}
\item[$^{(1)}$] With respect to the maximum of the optical
lightcurve, 
\item[$^{(2)}$] for V~Cam the second harmonic of the modelled
rotation period,
\item[$^{(3)}$] with respect to the null hypothesis (data are
Gaussian noise),
\item[$^{(4)}$] with respect to the local standard of rest,
\item[$^{(5)}$] order-of-magnitude estimate based on the Zeeman
hypothesis, and for a linewidth of 1\,\kms. This corresponds to
a maser at $3\,r_{\rm J}$ and a planetary magnetic dipole field
eight times stronger than Jupiter's (Wiesemeyer, 2008).
\end{list}
\end{table*}
\section{Results}
   \begin{figure*}
   \centering
        \resizebox{15cm}{!}{\includegraphics*{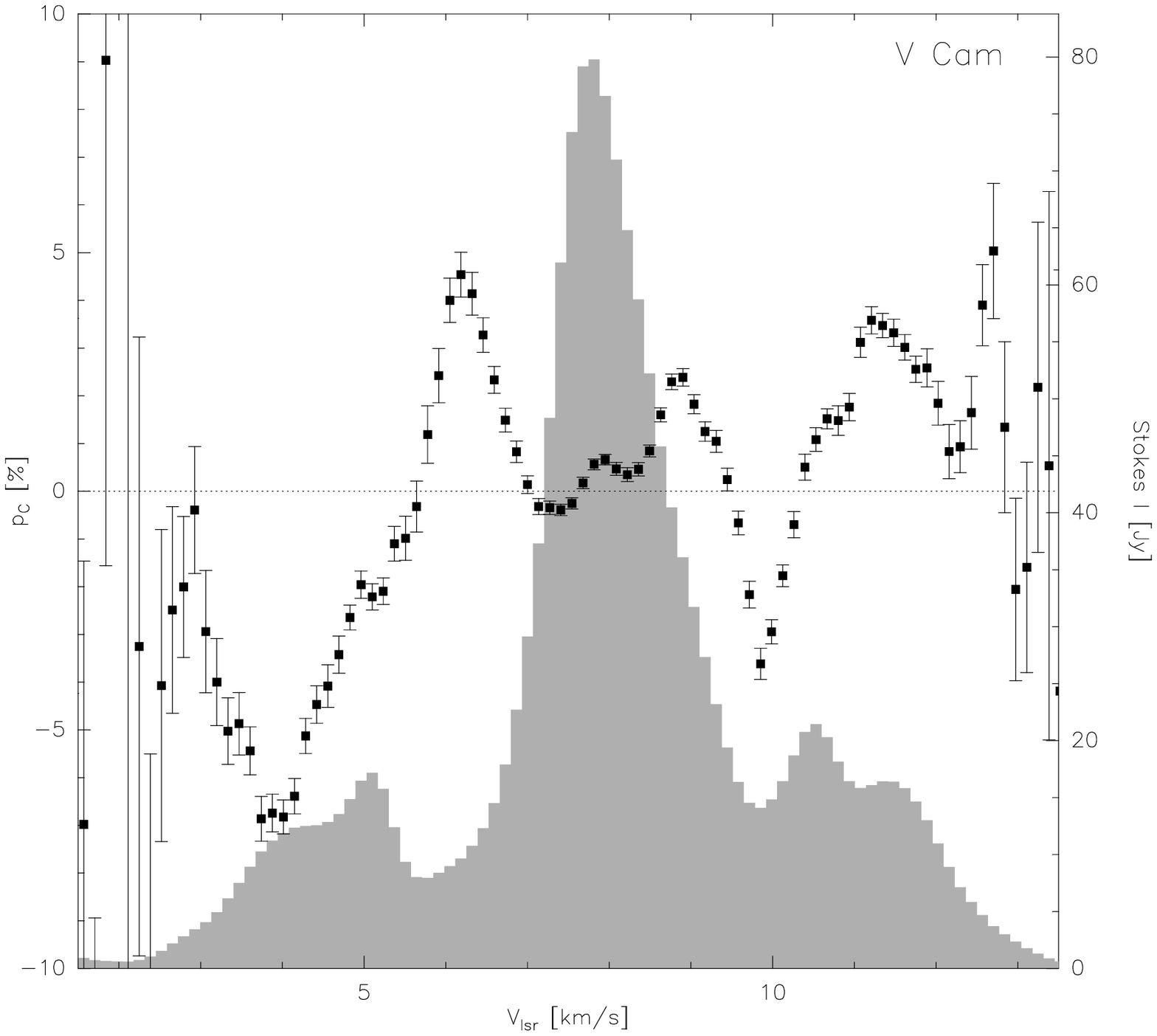}
                            \includegraphics*{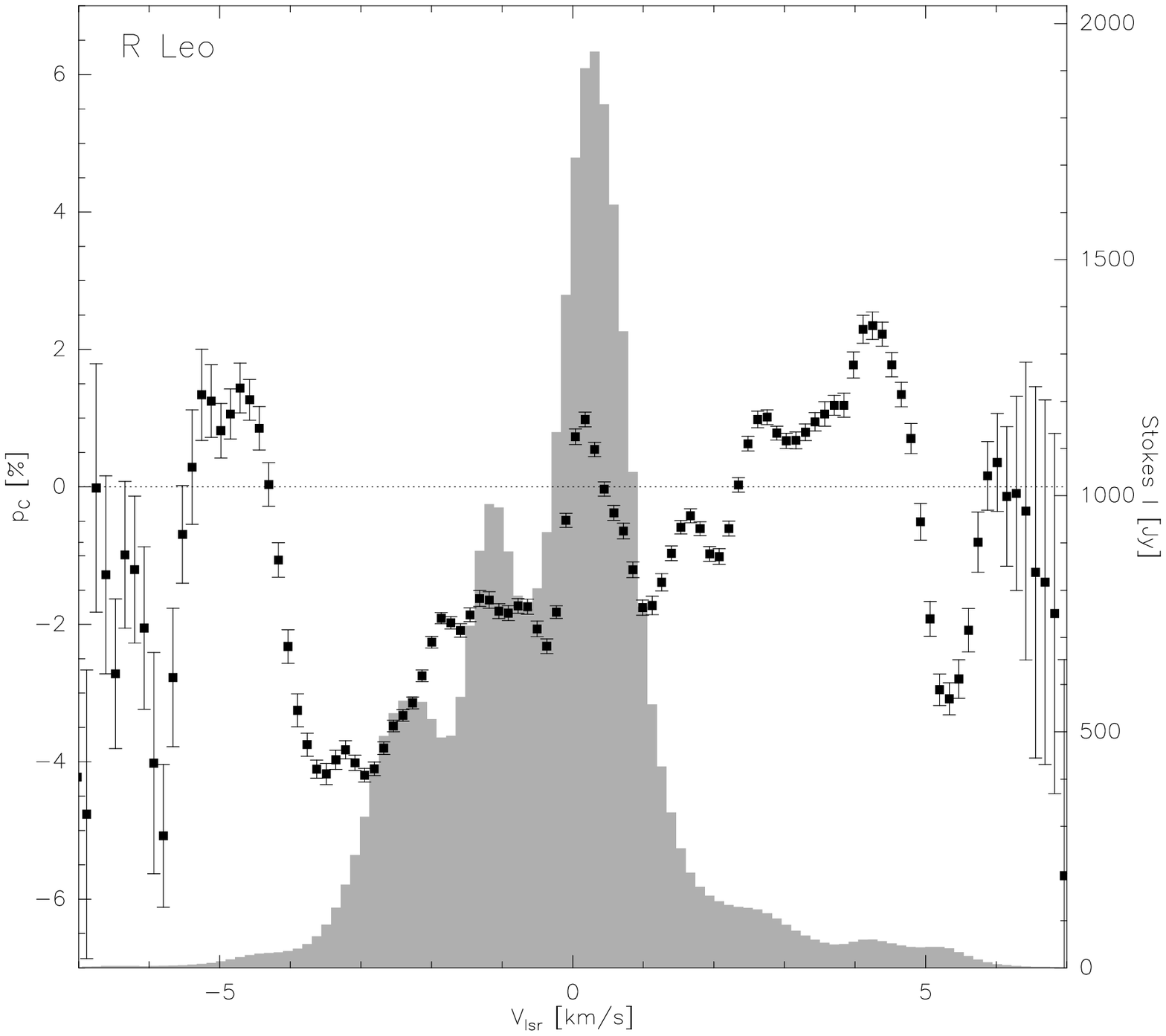}}
   \caption{Spectra of total power (Stokes I, i.e. polarized plus unpolarized flux density) and of fractional circular
            polarization. Grey-shaded histogram: Stokes I flux density in Jansky vs. {\vlsr} (the radial velocity
            with respect to the local standard of rest). Dots with errorbars: fractional circular polarization \pc, in \%
            with respect to Stokes I. Left: Mira-type star V~Cam. Right: Mira-type star R~Leo.}
   \end{figure*}
%
   \begin{figure*}
   \centering
        \resizebox{15cm}{!}{\includegraphics*{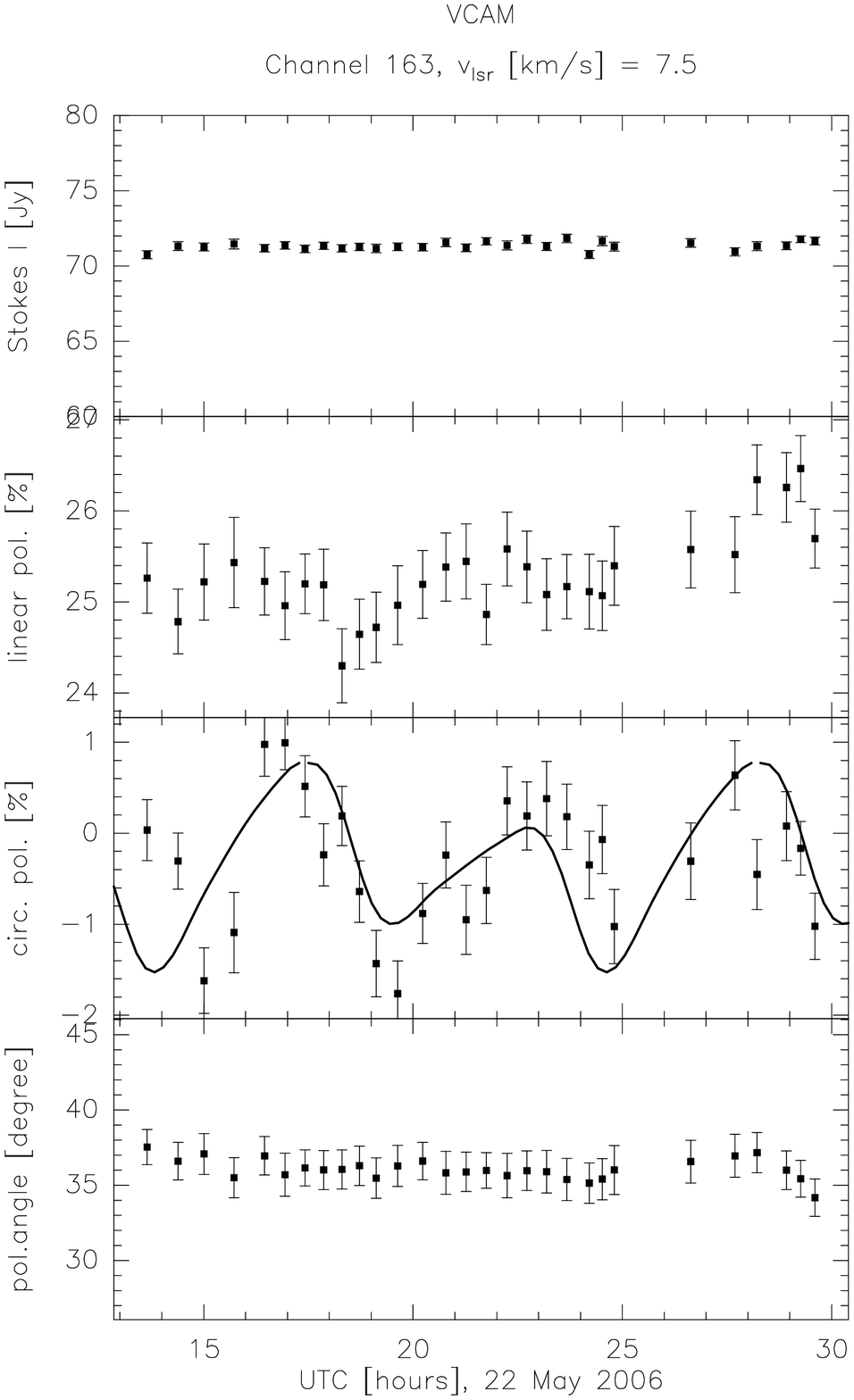}
                            \includegraphics*{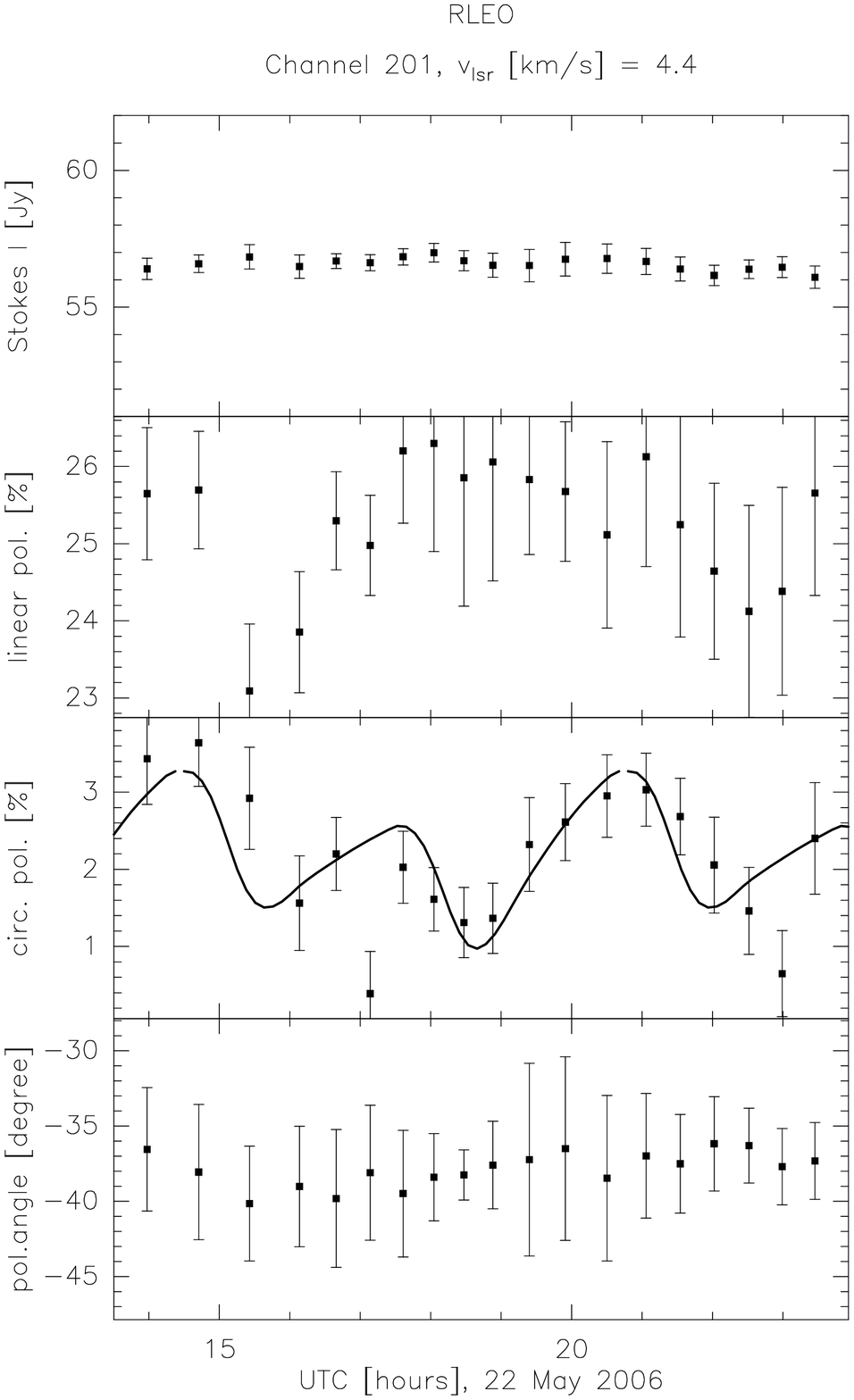}}
        \caption{Time series of polarization measurements. From top to bottom: Stokes I in Jansky, the
                 fractional polarizations {\pl} (linear) and {\pc} (circular), and the polarization angle (i.e. the position
                 angle of linear polarization, in degree E from N), for the SiO maser spots from V~Cam at
                 \vlsr\,$ = 7.5$\,\kms (left) and R~Leo at \vlsr\,$ = 4.4$\,\kms (right). The continuous
                 heavy lines in the plots for {\pc} show model results (Wiesemeyer, 2008) for a saturated maser in the
                 equatorial plane, at $3\,r_{\rm J}$ from the planet, with a magnetic dipole field of eight times Jupiter's
                 and misaligned with the rotation axis (in the sky plane) by $10^\circ$, and rotation periods of $10\fh 8$
                 (V~Cam, left) and $6\fh 3$ (R~Leo, right).}
   \end{figure*}
Only two stars, the Mira variables V~Cam and R~Leo (Fig.~1) passed the three tests and show clear evidence for a
quasi-periodic modulation of their fractional circular polarization $p_{\rm C}$ (Fig.~2), of peak-to-peak amplitude
2.8\,\% and 3.7\,\% for V\,Cam and R\,Leo, respectively (Tab.\,1).
The observations were made $\sim 40$ days after the optical maximum of V~Cam (pulsation period 522\,d) and $\sim 30$ days
after that of R~Leo (period 310\,d), in an expansion phase of the SiO maser shell. Within a $\sim 0.4$\,\kms narrow
range of radial velocities centred at $7.5$\,\kms and $4.4$\,\kms, we detect periods of $5\fh 4 \pm 0\fh 1$ and
$6\fh 3 \pm 0\fh 3$, for V~Cam and R~Leo, respectively (Tab.~1). The phenomenon is both rare and localized in velocity
space (and thus in circumstellar space, due to the velocity structure). The likelihood of misinterpretation of a time series
of Gaussian noise is $\le 4$\,\% and $\le 16$\,\% for V~Cam and R~Leo, respectively (Fig.~3, left panels). We note that the
Lomb periodograms of the Stokes U residuals do not show any significant periodic signal (Fig.~3, right panels).

The Monte Carlo tests confirm the significance of the results. For $10^5$ runs,
only 0.28\,\% of the Lomb periodograms mimic a periodic signal with $\le 4\,\%$ false-alarm probability (V~Cam),
respectively 2.9\,\% a periodic signal with $\le 16\,\%$ false-alarm probability (R~Leo). We therefore conclude that it
is unlikely (for V~Cam extremely unlikely) that random noise creates the observed fluctuations. If they were produced by
an oscillation of the telescope tracking, moving the polarized sidelobes across the line-of-sight towards the source, all
velocity channels would show fluctuations with the same period, which clearly is not observed and can therefore be safely
excluded.

Variations of the circular polarization of $4\,\%$ peak-to-peak amplitude were found again in August 2008 (Fig. 4), but
now within two and wider velocity intervals, from \vlsr\,$= -3.6$ to $-2.7$\,\kms and from $-0.5$ to 0.7\,\kms.
A meaningful Lomb analysis as in the case of the 2006 data was not possible for these data, because the observing interval
was too short to measure a full fluctuation period. The data (Fig.~4 left and middle) suggest a period of about 2\,hours with a guessed
uncertainty of about 0.5\,hours. It is interesting to note that in channels of relatively strong Stokes I emission, no
significant polarization fluctuation was found (Fig.~4, right), while now also the linear polarization seems to fluctuate at
about \vlsr\,$=0.6$\,\kms, where Stokes I is strong. Again, this rules out artifacts as the reason for the fluctuations. If
they were of instrumental origin, the whole spectral band would be affected.
   \begin{figure*}
   \centering
        \resizebox{14.2cm}{!}{\includegraphics*{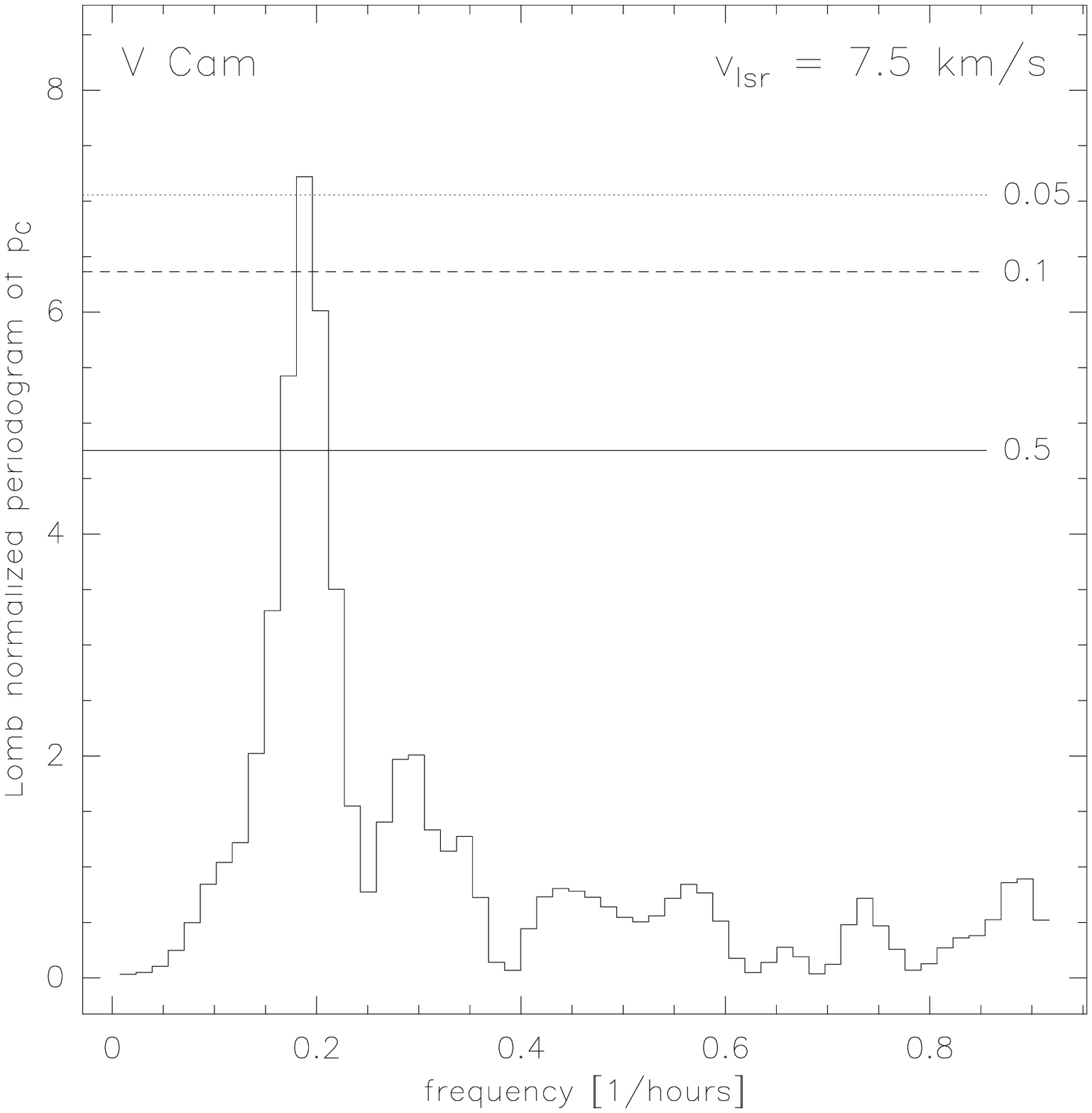}
                              \includegraphics*{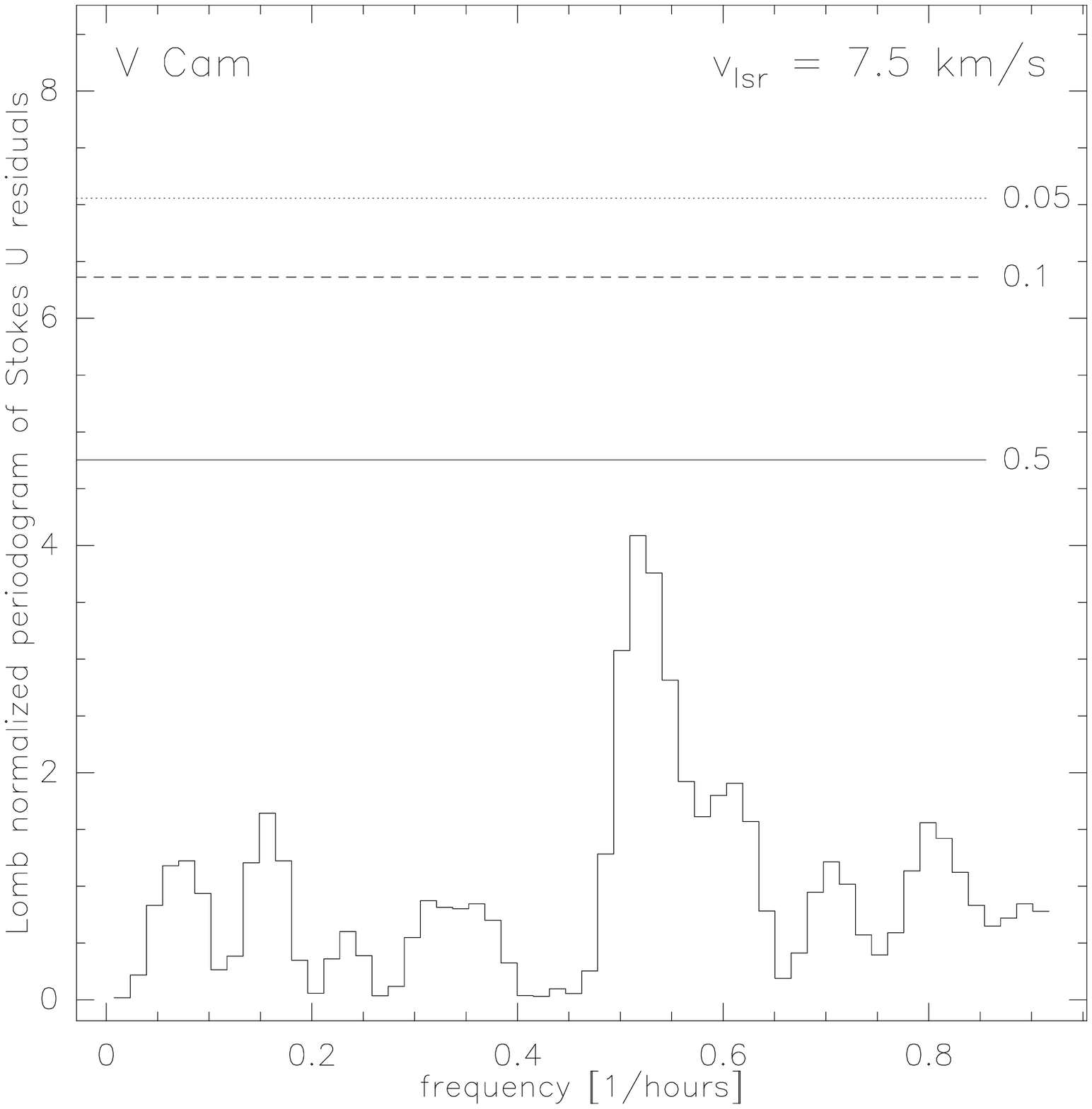}}
        \resizebox{14.2cm}{!}{\includegraphics*{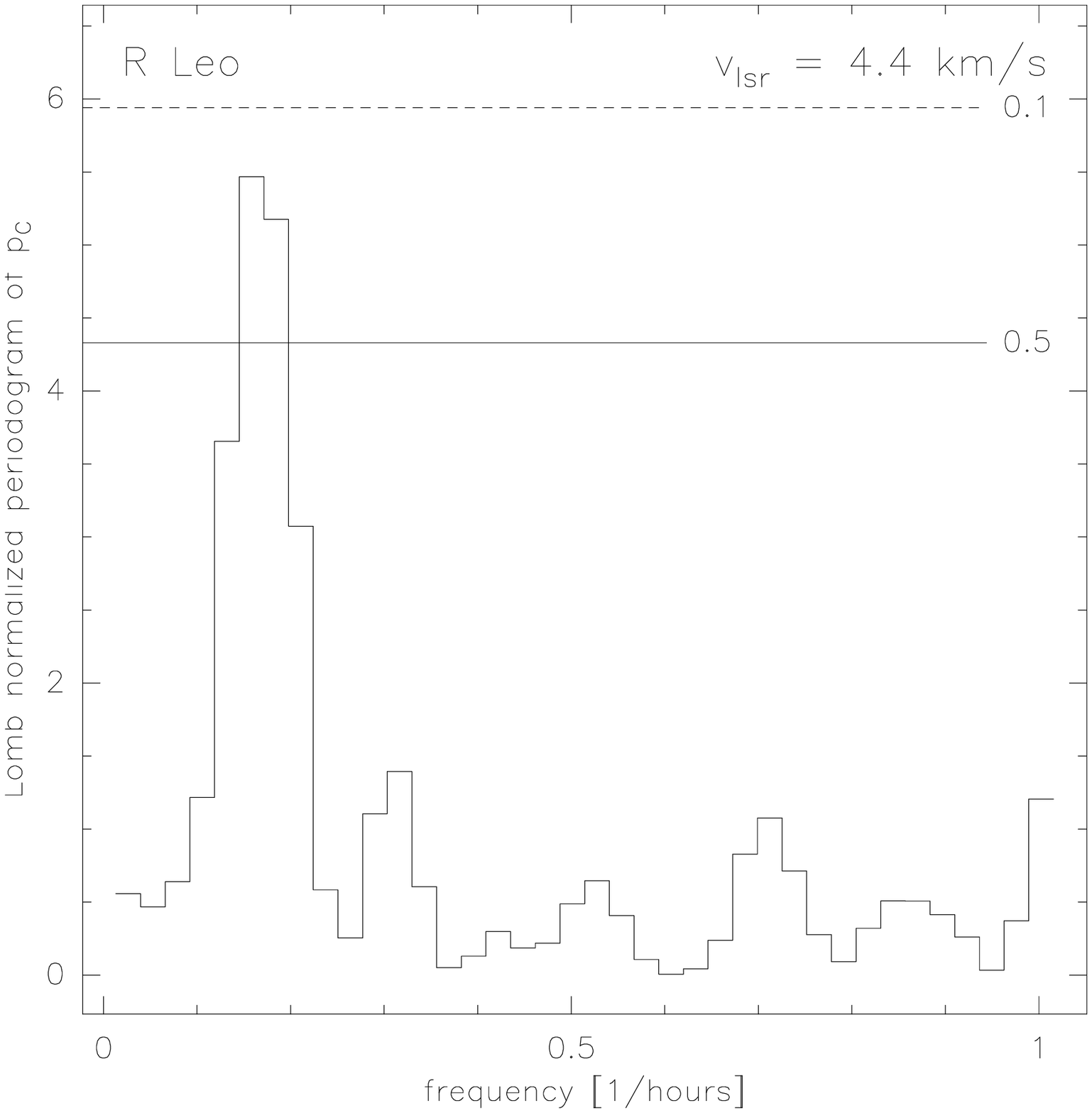}
                              \includegraphics*{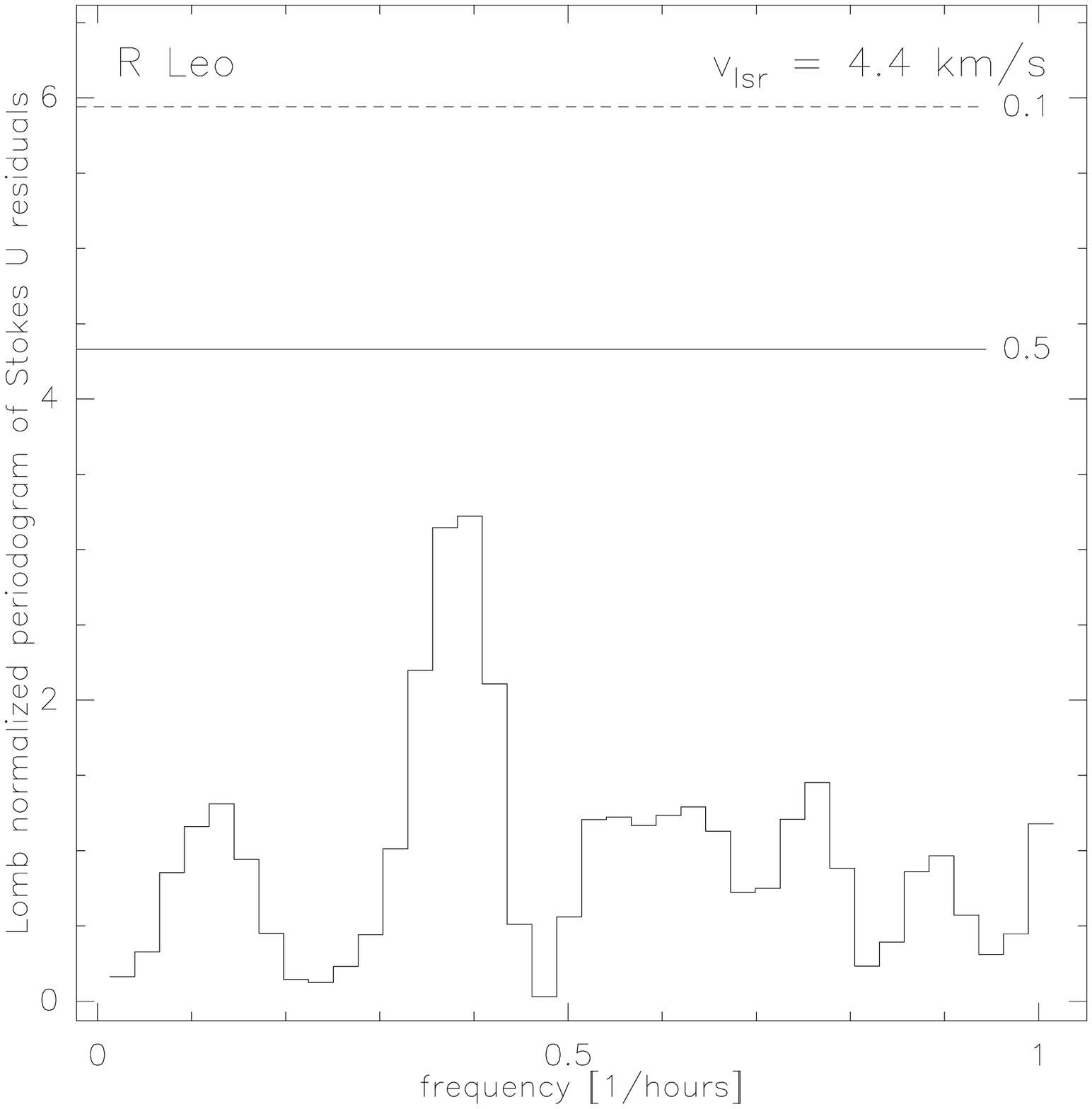}}
        \caption{Lomb periodograms of the polarization time series (estimates of the spectral power density
                 by least-square fits of their harmonic contents, Lomb, 1976), normalized by the variance of the samples.
                 Top left: Lomb periodogram of {\pc} from V~Cam at \vlsr\,$ = 7.5$\,\kms. Top right:
                 same for the residuals of Stokes U (normalized by Stokes I, after subtraction of the stationary linear
                 polarization). Bottom left and right: as top panels, but for R~Leo at \vlsr\,$=4.4$\,\kms. The horizontal
                 dashed lines indicate the spectral power density at which a periodic signal can be mimicked by Gaussian
                 noise with a probability given by the labels (''false alarm probability''). For a given source, the plot
                 scale for the Lomb periodograms of {\pc} and the Stokes U residuals is the same, to make the comparison of
                 the significance of the peaks easier.}
   \end{figure*}
   \begin{figure*}
   \centering
        \resizebox{19.5cm}{!}{\includegraphics*{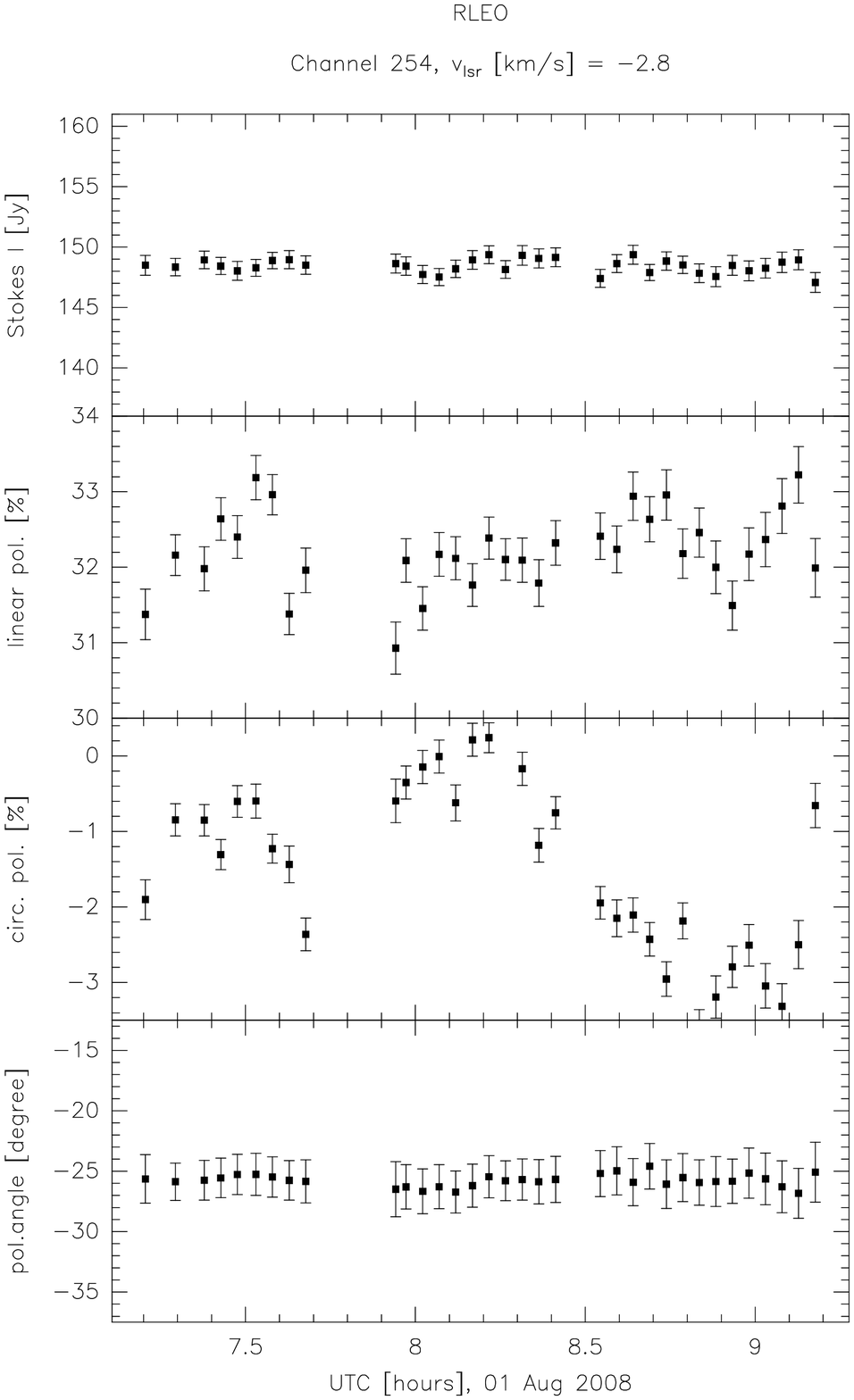}
                              \includegraphics*{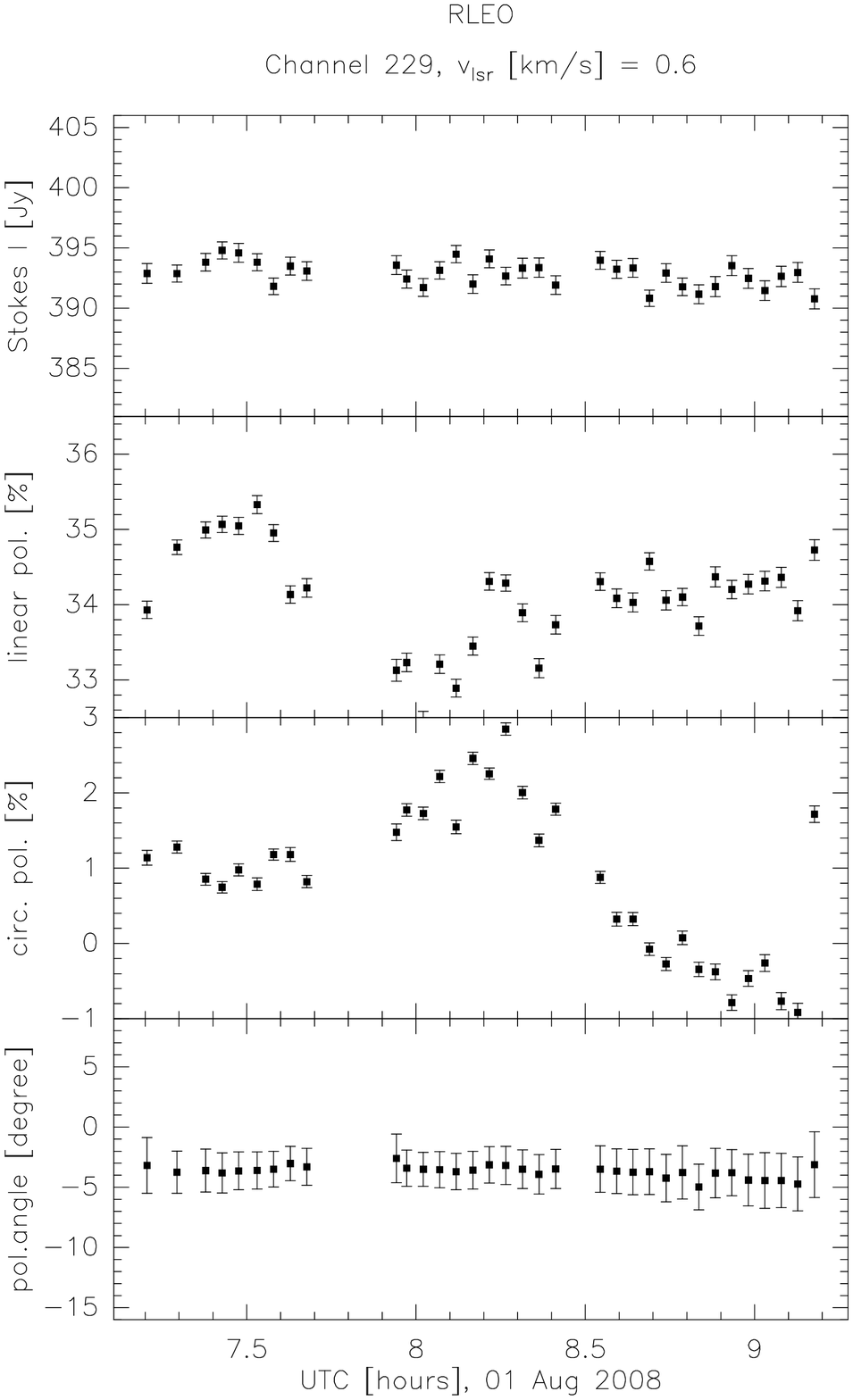}
                              \includegraphics{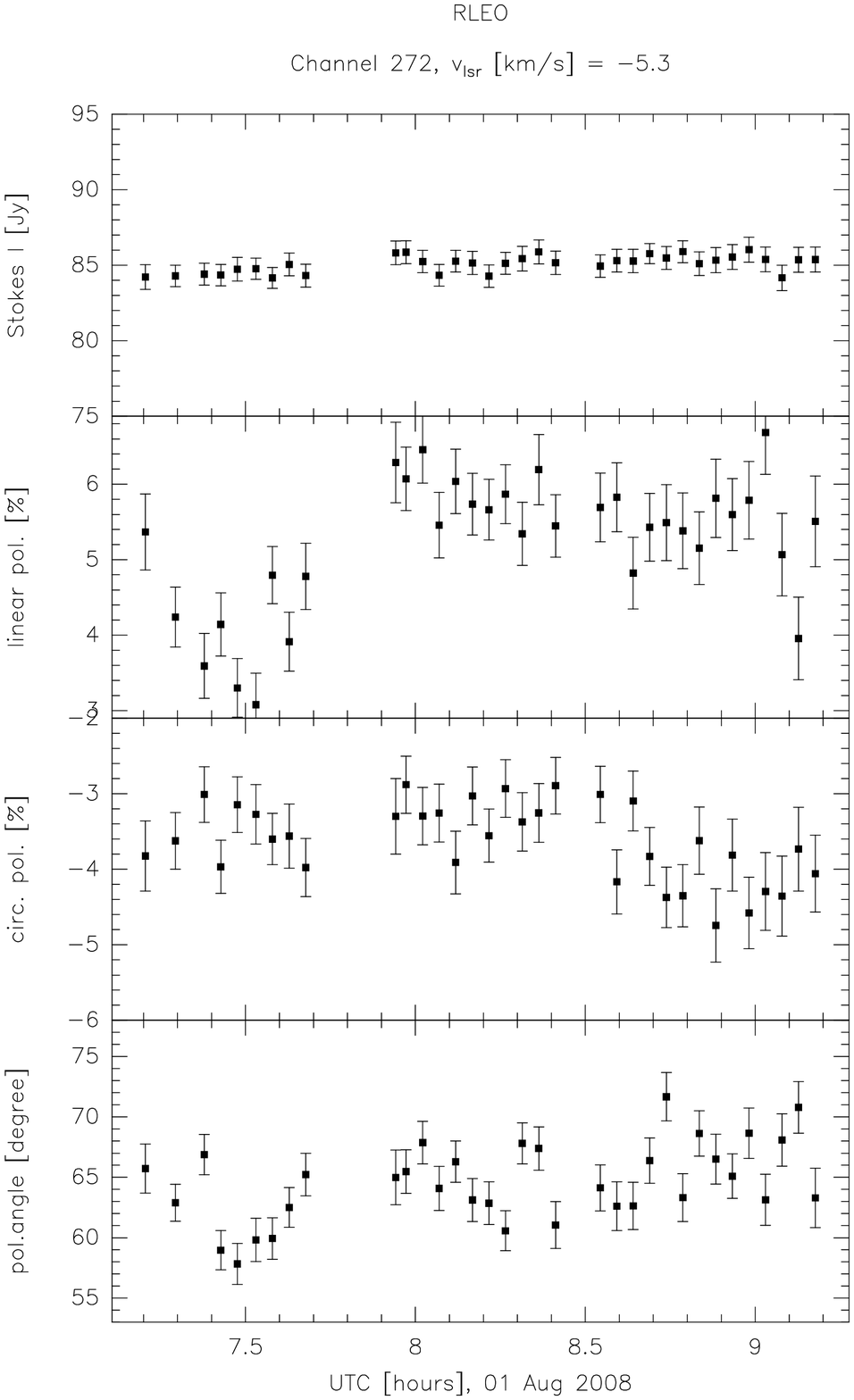}\hspace{0.5cm}}
	\caption{Same as Fig\,2, but for R\,Leo at three selected velocities (observations from 
	2008 August 01). Left: \vlsr\,$=-2.8$\,\kms. Middle : \vlsr\,=0.6\,\kms. Right: \vlsr\,$=-5.3$\,\kms}
   \end{figure*}
\section{Discussion}
The fractional circular polarization is a more reliable magnetic tracer than
linear polarization, especially for spatially unresolved observations: our telescope beam averages the polarization signal
from all maser spots at the same radial velocity, hence the orientation of the linear polarization is either parallel or
perpendicular to the maser shell (e.g. Diamond \& Kemball, 2003, Desmurs et al., 2000). This tends to decrease the
beam-averaged linear polarization, except if it is dominated by few, very strong maser spots close together (which may
explain the linear polarization fluctuations showing up in the August 2008 observations of R\,Leo, Fig.\,4, middle).
Furthermore, linear polarization can be entirely produced without magnetic fields by
anisotropic pumping (Western \& Watson, 1983, Asensio Ramos et al., 2005), though polarization angle swings by $90^\circ$
in the afore mentioned resolved observations are explained by changes of the magnetic field direction with respect to the
propagation direction of radiation in the maser (Goldreich et al., 1973). By contrast, circular polarization is
enhanced in magnetized maser spots, either directly via the Zeeman splitting (e.g. Watson \& Wyld, 2001), or
indirectly via birefringent conversion of linear into circular polarization due to changes of the magnetic field
topology along the maser slab (Wiebe \& Watson, 1998, for unsaturated masers), or due to maser saturation, rotating
the quantization axis from the direction of the magnetic field to that of the radiation propagation in the maser
(Nedoluha \& Watson, 1994). In the latter case, pump conditions would be required not only to vary in time but also
quasi-periodically, an event which we cannot exclude but seems unlikely on short time scales. In turn, this would
produce a varying Stokes I flux density, which is not observed. We therefore suggest that the intraday fluctuations
of {\pc} are of magnetic origin. They are quasi-periodical, not stochastic, and rare and only at
one velocity, which means that an ordered magnetic field structure with smooth gradients in time
and space is locally maintained within the dense AGB winds of R\,Leo and V\,Cam.
\subsection{A case for precessing Jovian magnetic fields ?}
The magnetopheres of Jovian planets have already been proposed to explain several features of SiO
maser polarization (Struck et al., 2002, 2004), thus providing a local explanation for the
strong magnetism in the atmosphere of AGB stars. If the magnetic fields in which SiO masers are embedded, of typically
10\,G, were global, one would assume, at the stellar surface, a magnetic field of about 100 to
1000\,G, depending on assumptions on the gradient of the magnetic flux density. Such a field could 
be generated by a dynamo between the rotating core and the stellar convection zone (Blackman et
al., 2001). However, we now have to explain (see item 3 below) why Mira stars were not detected
(Haisch et al., 1991) by the Rosat all-sky soft X-ray survey, which is sensitivity-limited to
X-ray luminosities above typically $2.4\times 10^{29}$\,erg\,s$^{-1}$. As a matter of fact,
Kastner \& Soker (2004a) find that the X-ray luminosity of the single Mira stars TX\,Cam and
T\,Cas is below the threshold expected for a dynamically important magnetic field. Their conclusion relies
on the assumption that the X-ray luminosity equals the kinetic luminosity of the stellar wind powered by 
magnetic fields. Since there may be ways to dissipate magnetic energy other than by X-ray emission from accelerated
particles, this is still an open issue.

We now summarize the arguments supporting the plausible planetary origin of our detections.
\begin{enumerate}
\item The rareness of the phenomenon can be explained by the number of Jovian planets around AGB stars, and by the
requirement that the gain region of the maser needs to cross the magnetosphere where it is dominated by the dipole
component. At present we cannot distinguish which explanation prevails - the possible rareness of Jovian planets in
AGB atmospheres due to infall of planets into the star (Villaver \& Livio, 2007, Nordhaus \& Blackman, 2006), or the
unfavourable location of the SiO maser with respect to the precessing magnetosphere.
\item Both in V~Cam and R~Leo, the signature is statistically significant only at one velocity. We interprete this
velocity as the current line-of-sight velocity of a planet on its Keplerian orbit. 
\item Dedicated models (Wiesemeyer, 2008) of saturated SiO maser emission in a precessing magnetic dipole field show that
a dipole moment eight times stronger than that of Jupiter's magnetic field is sufficient to provide the $\sim 1$\,G
magnetic flux density at $3\,r_{\rm J}$ distance, required to explain the peak-to-peak variation of {\pc}
(Tab.~1). This reconciles maser polarization and the lack of an X-ray luminosity expected for a stellar corona. Saturated
maser action is not required here, if the weak circular polarization of unsaturated masers is enhanced by the combined
effect of anisotropic pumping and magnetorotation (Wiesemeyer, 2008).
\item The fluctuations are quasi-periodic, not stochastic. The fluctuation periods derived from the Lomb periodograms of
the {\pc} fluctuations in R~Leo and V~Cam can be modelled as the first and second harmonic, respectively, of the planetary
rotation period which is $6\fh 3$ and $10\fh 8$, respectively (cf. Jupiter's rotation period of $9\fh 9$, and Saturn's
of $10\fh 5$). A misalignment of the dipole axis with the planetary rotation axis (10$^\circ$ in the models, $9\fdg 6$ for
Jupiter) naturally explains both the amplitude and the period of the fluctuation, due to the ordered 
structure of the magnetic field. This implies that the magnetosphere is
protected by a Chapman-Ferraro type magnetopause, where ram pressure equals magnetic pressure, at about $10\,r_{\rm J}$ (Jupiter radii)
substellar distance (according to Struck-Marcell, 1988). To date it is still uncertain why
our R\,Leo data from 2008 yield a different period. This may be partly due to the incompleteness of
the data, partly due to the proper motion of the maser spots, which may cross $10\,r_{\rm J}$
in 10\,hours and thus modify the pattern expected for a stationary fluctuation. Any intrinsic origin (changes of the 
angular momentum of accretion onto the planet, or of the spin of the latter) would be most difficult to model. Likewise,
it is still unclear why in 2008 polarization fluctuations appeared in two well separated velocity ranges.
\item If the pseudo-periods of the fluctuations were only due to the proper motion ($\sim 10$\,\kms) of SiO maser
spots (with $\sim 0.1$\,AU diameter) across a quasi-stationary magnetic perturbation, this would result in periods which
are two orders of magnitude longer than the observed ones. In turn, this means that the magnetic field diffuses through
the gas of the maser slab, i.e. the ambipolar diffusion speed exceeds the maser velocity. A Jovian magnetic field both
provides the required strength and curvature radius of the field lines (Struck et al., 2002), unlike a magnetic field of
stellar origin. 
\item Spatially resolved observations (Cotton et al., 2008) of the $v=1, J=1-0$ SiO masers in R~Leo show, at
$\sim 4 - 7.9$\,\kms radial velocity (with the largest redshifted velocity closest to the star), a remarkable elongated
feature suggestive of a planetary wake flow as modelled by Struck et al. (2004). The corresponding feature in the
$v=2,J=1-0$ transition is even double-lobed. Similar features were found by Cotton et al. (2004) and Soria-Ruiz et al.
(2007) at various circumstellar positions and different epochs.
\end{enumerate}
\subsection{Kinematical and astrometric evidence for a planet orbiting around R\,Leo}
We deliberately consider here the presence of elongated features (see item 6 above) as the possible signature of a
planetary wake although explanations involving the theory of maser formation cannot be ruled out. In the frame of this
working hypothesis we now will use the available kinematic and astrometric information to illustrate the potential
interest of combining SiO polarization monitoring observations with contemporaneous VLBI imaging of SiO masers.
The VLBA
observations of Cotton et al. (2008) were in September 2004, 638\,days after similar observations by Soria-Ruiz
et al. (2007) in December 2002, while we observed in May 2006, 624\,days after Cotton et al. (2008). Remarkably, the
$v=1, J=1-0$ map of Soria-Ruiz et al. shows, at about \vlsr=\,$-7$\,\kms (with respect to the local standard of rest, hereafter
LSR), a similar elongated maser feature pointing radially away from the star towards the South-East, and another one 
towards North-East, with velocities peaking at \vlsr\,$=2.5$\,\kms and $-2$\,\kms.
This suggestion is confirmed by inspection of the VLBA map taken in August 2001 by Cotton et al. (2004), again showing a
quasi-radial alignment of maser spots, now towards the north. The repeated occurence of such features, systematically
rotated counter-clockwise from North, strengthens the suggestion of a wake flow associated with the same planet.
However, the combined use of $v=1,J=2-1$ polarization monitoring with the $v=1,J=1-0$ VLBI astrometry raises new
questions. Soria-Ruiz et al. (2007) show that in R\,Leo the former maser spots do not coincide with those corresponding
to the latter transition. The characteristic ring radii of the respective maser shells are 29.2\,mas and 33.8\,mas,
for the $J=1-0$ and $J=2-1$ transitions, whereas collisional and radiative pumping models (see Soria-Ruiz et al. for
further references) suggest that for our purposes the maser spots of these transitions should be sufficiently close to each
other. With this inconsistency in mind a lower limit to the astrometric error in the comparison of our polarization data
with the VLBA positions can therefore be estimated from the difference between the ring radii, 4.6\,mas or 0.5\,AU
(for 113.5\,pc distance, Fedele et al., 2007, further references therein). The actual astrometric error may be higher,
because the absolute astrometry of these VLBI data from different epochs is far less well constrained than the relative
astrometry for a single observing date. Despite this limitation we use here the available combined kinematics and 
astrometric information to constrain the planetary orbit. The velocity in the wake flows is usually decelerated, the
maximal blue- respectively redhifted velocity should therefore be closest to the planet. A sine fit (Fig.~5, top) yields a
period of 5.2\,years and a stellar velocity of \vlsr\,$= 1.7\,$\kms, but more observations are needed in order to assess
the significance of the fit. Our 2008 data are best described by the velocity of the more blueshifted polarization
fluctuation. The fitted stellar velocity is between the value derived by Bujarrabal et al. (1989) from thermal
molecular lines in the circumstellar envelope ($-0.5$\,\kms) and the catalogue value of
$13.4$\,\kms (GCRV, Wilson, 1953). Using the method discribed in Appendix\,A, this velocity fit
allows us to determine the radius of the planetary orbit to 23.8\,mas which corresponds to 2.7\,AU,
at a distance of 113.5\,pc. The inclination of the orbital plane with respect to the plane of the sky is
$i=34^\circ$. The result of the least-square fit to the available astrometric information is shown in Fig.~5 (bottom).
The linear polarization measured by us is predominantly parallel to the
radius vector calculated from the model fit. Due to uncertainties in our method and in the astrometric 
positions of the VLBI spot maps (no absolute uncertainties are derived in the works of
Cotton et al. and Soria-Ruiz et al.), it is not too surprising to observe a discrepancy between
the observed and modelled positions, especially for the 2002 Dec 07 VLBA observation (Soria-Ruiz,
2007). 

As for the August 2008 data in R\,Leo which show polarization fluctuations at two different velocities,
contemporaneous VLBI observations would have been required to perhaps further suggest that the second
velocity interval could be due to a wake flow and therefore to another planet. However, it seems too
speculative at this stage to suggest that each elongated VLBI maser feature could be associated with a
planet while jet-like features could as well be interpreted as ejections of matter (see e.g. Cotton et al.,
2008). Our fit leads to a stellar mass estimate of $0.7\,M_{\sun}$, below, but not unreasonably far from
the value inferred from measurements of stellar parameters by infrared K-band interferometry (Fedele et al., 2005),
namely $1.0-1.2\,M_{\sun}$, for an equivalent non-pulsating star of the same luminosity. The discrepancy
can either be due to observational error or model uncertainties. As a matter of fact, the R\,Leo maps of
Cotton et al. (2008) mentioned above show several elongated maser features, and it is unlikely that each is
associated with a planet (Cotton et al. rather suggest jet-like ejections of matter, see section 4.3.6 below).
The conclusion that the polarization fluctuations observed at two velocities in R\,Leo in 2008 are due to the 
presence of two planets would therefore be speculative, even if contemporaneous VLBI data were at hand.
   \begin{figure}
   \centering
        \resizebox{12cm}{!}{\includegraphics{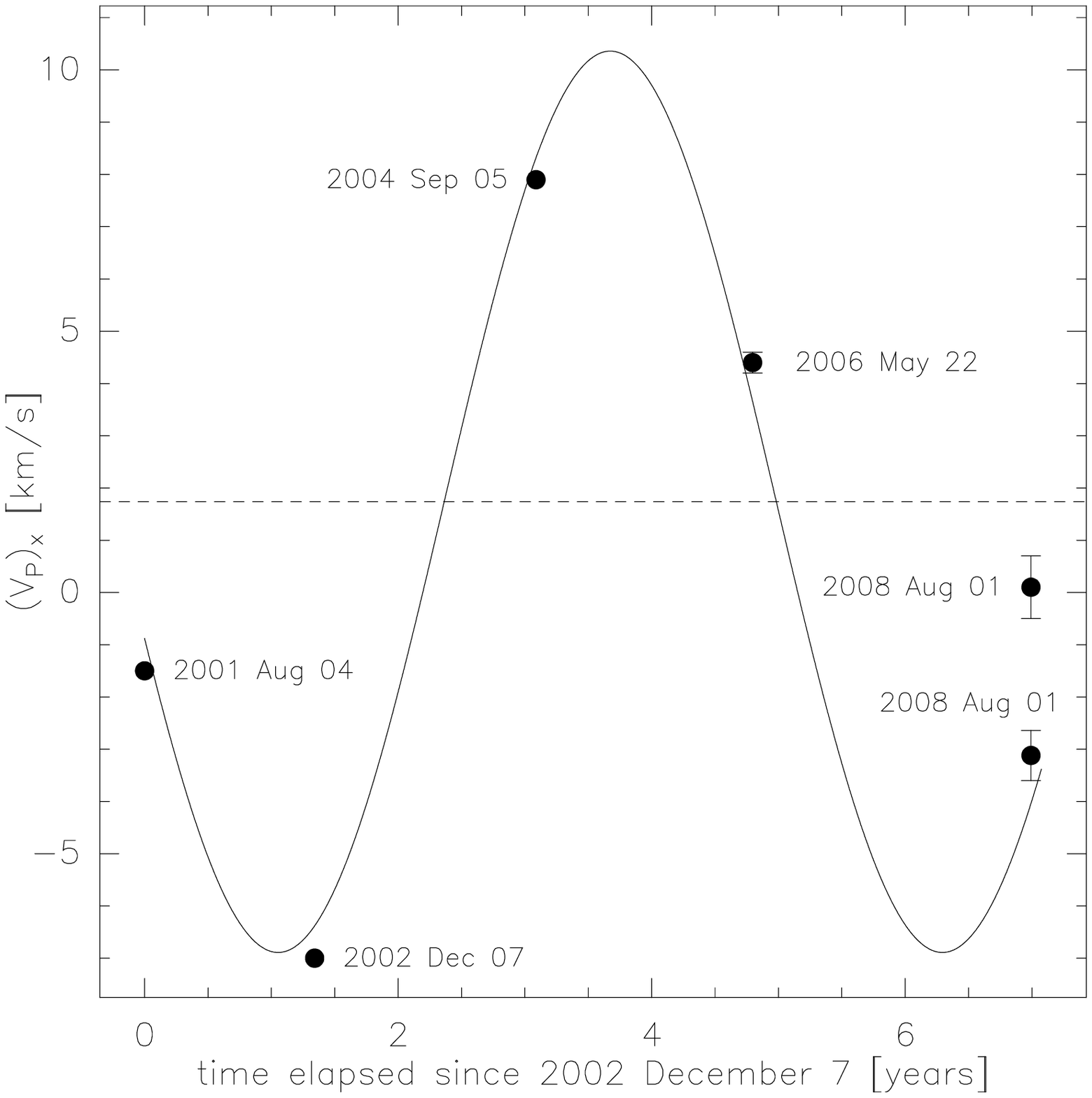} \hspace{5cm}}
        \resizebox{12cm}{!}{\includegraphics{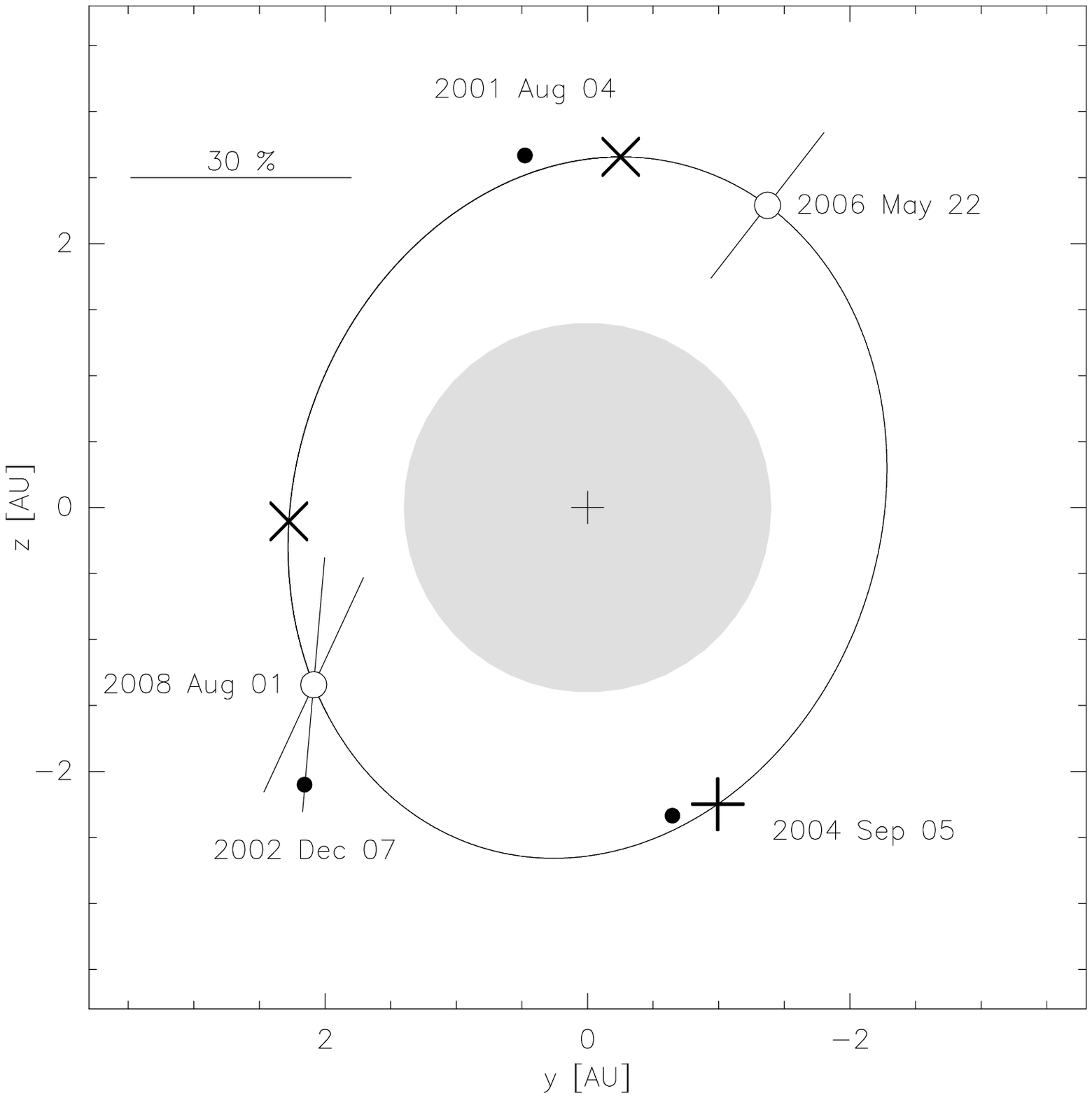} \hspace{5cm}}
        \caption{Top: Keplerian velocity fit of maser features associated with the suggested planet
	         (LSR velocities at the tip of the planetary wake flow, and velocities of polarization fluctuations).
                 Time is in years since 2001 Aug. 04 (observations of Cotton et al., 2004). Other references: Soria-Ruiz et 
                 al., 2007 (2002 Dec. 07), Cotton et al., 2008 (2004 Sep. 05), this work (2006 May 22 and 2008 Aug. 01).
                 Bottom: available astrometric information and least-square fit of the derived planetary orbit
                 (Appendix~A), for $d=113.5\,$pc stellar distance. VLBI positions are indicated by black dots. The
                 planetary orbit is indicated by the black line, crosses mark the fitted position at the VLBI epochs.
                 Empty circles give the positions of the planet at the epochs of our polarization measurements.
                 The linear polarizations measured by us and their position angles are shown as bars (the bar in the top
                 left corner indicates a fractional linear polarization of 30\,\%). The grey-shaded disk indicates the
                 measured $\lambda\,2.2\,\mu$m uniform-disk diameter at stellar phase 0.08 (in April 2001, Fedele et al.,
                 2005).}
   \end{figure}
\subsection{Are alternative scenarios viable ?}
Boldly extrapolating the rich magnetic phenomena in the solar system to late-type stellar systems, alternative
scenarios have to be seriously examined. One might suggest magnetic activity due to stellar companions, coronal loops,
magnetic clouds or Parker instabilities in shocks forming in response to stellar pulsations. 
\subsubsection{Binarity}
Here the mass transfer from the AGB star to a close companion generates flares and related magnetic fluctuations, which
are not necessarily located in the atmosphere of the former (e.g. the Mira AB system, Kastner \& Soker, 2004b). However,
neither V~Cam nor R~Leo are known for having stellar companions, otherwise their X-ray luminosities would be higher. As a
matter of fact, Tatebe et al. (2006, IR interferometry at $11.5\,\mu$m) do not see any obvious asymmetry in the envelope
of R\,Leo that could hint at a companion (see also Gatewood, 1992).

The case of the 2003 soft X-ray outburst in the Mira AB system is, interestingly, not related to the binarity of the system.
The spatially resolved Chandra observations (Karovska et al., 2005) clearly show that it occured within the envelope of Mira A,
and it would be worthwhile to estimate whether it was triggered by burst-like accretion onto a Jovian planet, as suggested
by Struck et al. (2004). As a matter of fact, the position of the X-ray flare may coincide with a linear SiO feature seen by
Cotton et al., (2004).
\subsubsection{Coronal loops}
Coronal loops are generated by the differential rotation in the sub-photosphere, together with the dynamo (Blackman et al.,
2005) at the interface with the core. If the convection zone extends to at least 80\,\% of the stellar radius
(Holzwarth \& Sch{\"u}ssler, 2001), magnetic flux loops are trapped there, because magnetic stresses at the tips of the
loops act against convective buoyancy. This also explains the lack of a coronal X-ray luminosity, and why the magnetic
fluctuations are rare. One may suggest coronal loops to occasionally extend to the site of the SiO masers. If there were coronal flux
loops, the timescale for magnetic fluctuations would be too long, though. This can be demonstrated in
the following way: Assuming (by analogy with the sun) that the size of a loop corresponds to that of a super-granulation
cell, this yields 0.2 AU, for 1\,AU stellar radius and 400 convection cells in the stellar sub-photosphere (Schwarzschild,
1975), in contrast to $2 \times 10^6$ for the sun. Interferometric K-band size measurements (Fedele et al., 2005)
yield a Rosseland radius of 1.6\,AU at phase 0.08, close to the brightness maximum in April 2001 (cf. our measurements
at phase 0.08 after the brightness maximum in 2006). Our orbital radius estimate of $2.7$\,AU thus infers that the maser
is 1.1\,AU above the stellar photosphere. Assuming a photospheric rotation speed of $\sim 1$\,\kms (a higher rotation
speed is only possible for AGB stars with at least a brown dwarf companion, Soker 2006), a corotating flux loop would need
$\sim 100$ days to cross a maser spot of 0.1\,AU diameter. This is two orders of magnitude above our fluctuation periods,
which makes it highly unlikely that coronal loops can generate the observed hour-scale magnetic fluctuations. The 
scenario involving trapped magnetic flux loops also demonstrates that from the weak X-ray luminosities one cannot infer the
absence of an efficient stellar dynamo - the magnetically accelerated particles simply remain within the convective envelope
and do therefore not show in X-ray observations.
\subsubsection{Magnetic clouds}
As coronal loops, magnetic clouds originate from the magnetic activity on the surface of the stellar photosphere. If coronal
loops are absent, due to a lack of the latter, this also holds for magnetic clouds of the type encountered in the solar
wind (e.g. Bothmer \& Schwenn, 1998). Should magnetic clouds traverse the extended stellar atmosphere, we can provide a
rough estimate of the time scale associated with the corresponding magnetic fluctuations. The size of magnetic clouds in
the solar system is comparable to that of SiO masers (0.1 AU), and their magnetic field vector also rotates on a timescale
of $\sim 10$ hours. However, solar system magnetic clouds expand with the speed ($\sim 400$\,\kms) of the magnetically
driven solar wind. Since the latter is an order of magnitude above that of an AGB wind and SiO maser proper motions, the
timescale does not fit - the resulting polarization would vary too slowly (especially if the magnetic cloud is decelerated
by the dense ambient gas). 
\subsubsection{Parker instabilities}
In order to produce polarized maser features of the observed lifetime (months to years, Glenn et al., 2003), the latter
needs to be shorter than the ambipolar diffusion timescale, i.e. the magnetic flux has to be coupled to the masing matter.
The fragments forming as a consequence of Parker instabilities (Hartquist \& Dyson, 1997) fulfil this requirement under
the typical conditions in the extended atmosphere of AGB stars. Since Parker instabilities arise from a local indentation
orthogonal to the field lines, some short-term variations of the magnetic field component along the line of sight may be
expected if the ambipolar diffusion timescale of these fluctuations exceeds that of the maser lifetime and proper motion.
The resulting variability of {\pc} should globally affect all maser spots of an SiO star if they commonly arise from
fragments formed by this mechanism, in contradiction with our observations. Furthermore, the mass of fragments forming in
response to Parker instabilities scales with magnetic pressure, and the strongest fluctuations would be expected from the
strongest masers, which has not been observed (cf. Figs.\,2 and 4). 
\subsubsection{Alfv\'en waves}
Our discussion would not be complete without considering Alfv\'en waves (Alfv\'en, 1942, see e.g. Sturrock, 1994)
as a possible origin of rapid magnetic fluctuations. However, our knowledge of hydromagnetic wave propagation in the dense AGB 
atmospheres is poor. The relevant time scale here would be
\begin{equation}
t_{\rm A} = D/v_{\rm A}
\end{equation}
where $D$ is the diameter of a maser slab, and $v_{\rm A}$ is the Alfv\'en speed for the propagation of a hydromagnetic wave, given
by (cgs units)
\begin{equation}
v_{\rm A} = \frac{B}{\sqrt{4\pi \rho}}
\end{equation}
where in the weak-coupling regime $\rho$ is the ion mass density, and in the strong coupling regime (i.e. when collisions between ions
and neutrals are important) the total mass density of neutral and ionized particles (Zweibel, 1989). Which regime holds in the
atmosphere of AGB stars is uncertain. If the ambipolar diffusion speed is close to the velocity of the AGB wind, as estimated by Soker 
(2006) and Hartquist \& Dyson (1997), strong coupling holds, and we estimate the Alfv\'en speed to
\begin{equation}
v_{\rm A} = \frac{10\,{\rm G}}{\sqrt{4\pi\,2.10^{-14}\, {\rm g\, cm}^{-3} }} \sim 200\, {\rm km s}^{-1}\,.
\end{equation}
for a 10\,G magnetic field and a hydrogen number density of $10^{10}\,{\rm cm}^{-3}$. Thus, the timescale for an Alfv\'en wave 
crossing a maser slab of 0.1\,AU diameter is about 20\,hours, not unreasonably far from the timescales given in Table~1. However,
we would like to stress that the parameters used for the estimate of both the ambipolar diffusion and Alfv\'en speeds are
much uncertain.
\subsubsection{A precessing jet ?}
The role of a planetary or brown-dwarf companion in the shaping of circumstellar envelopes has been modelled by Nordhaus \& Blackman
(2006). Since in their scenarios the companion is already engulfed in the convective envelope of the star, it cannot be detected anymore
by polarimetry. The deposit of the planetary angular momentum leads to a spin-up of the star and by consequence to the ejection of matter
by equatorial mass loss or, if assisted by the stellar magnetic field, to a bipolar jet. Is it conceivable that we observed rather
the polarization signature due to the precession of such a jet ? If the precession is caused by the misalignment between the jet axis and
the stellar rotation axis, we can estimate the timescale of the phenomenon as in the case of coronal loops (section 4.3.2.) except that 
we have to account for the spin-up of the star. Nordhaus \& Blackman showed that an infalling brown dwarf of $0.05\,M_{\sun}$ may increase
the stellar rotation rate by an order of magnitude. The fluctuation of the circular maser polarization due to a crossing jet would therefore
occur on a timescale of at least 10~days if the cross-sectional area of the jet perpendicular to its axis is smaller than the size of the
maser spot, or longer otherwise. This is still one to two orders of magnitude above the fluctuations observed by us. Furthermore, to leave
a signature only in a few SiO masers the opening angle of the jet cannot be too large.
\section{Conclusions}
In summary, in our opinion none of the non-planetary alternatives satisfactorily explains the present observations of
R\,Leo and V\,Cam, while a single scenario, namely precessing Jovian-type magnetospheres, are consistent with all of them.
Since the discovery of a planet around a star in the post-red giant phase (Silvotti et al., 2007), we know that planets can survive
in such an environment. Whether a given planet survives the AGB phase of its parent star depends on several factors
(Villaver \& Livio, 2007). Evolution of the orbit is the main one - depending on its distance from the star, the planet is
either dragged inwards by tidal friction and evaporated, or repelled due to stellar mass loss, while for Jupiter-sized
planets the gain of mass due to Roche flow from the star or accretion from the stellar wind barely affect their orbits.
If the suggested planet around R~Leo, one of the best studied Mira stars, of solar mass and at 113.5\,pc distance (Fedele
et al., 2005), does not spiral in and evaporate during the remainder of the upper AGB phase, it will survive the subsequent 
planetary nebula stage against evaporation (Villaver \& Livio, 2007) only if it has at least two Jupiter masses and orbits
at least at $\sim 3$\,AU. We note that the fluctuating SiO maser feature in R\,Leo has an orbital radius of 2.7\,AU,
close to the critical distance. We have demonstrated that the orbital elements of its host planet can be determined
by measuring the velocity of the fluctuation and the position of the corresponding maser spot. The orbital period of the
planet is 5.2\,years, which provides an estimate of the stellar mass of $0.7\,M_{\sun}$. This agrees with mass
estimates found in the literature, of about $1\,M_{\sun}$ for R\,Leo. On the long term, our method has the potential to
reveal, together with an independent estimate of the stellar mass, sub-Keplerian orbital motion and thus migration of the
planet through the atmosphere towards the star. This requires correspondingly accurate observations. The situation may be
complicated by the presence of several planets, whose orbits could be constrained by our method provided that the
polarization fluctuations are not affected by a spectral blend of independent features at the same radial velocity
and that the VLBI observations are free from a spatial blend of independent features along the same line of sight.
The estimated mass loss of R\,Leo, $\sim 10^{-7}\,M_{\sun}/{\rm y}$, is too weak to measure orbit evolution within a
reasonable amount of time ($\sim 3\,\mu$as in 1000 years). The situation may change if the opposite effect due to tidal
friction, neglected in this estimate, is by several orders of magnitude more important than that of stellar mass loss.

Without further observational evidence at hand, one can only speculate on whether the suggested Jovian magnetic fields are
decisive for evaporation or survival of the planets. They suppress turbulent viscosity and therefore the tidal friction
induced by the latter (Zahn, 1977), and also lead to spin angular momentum loss due to magnetic braking. This contrasts
with scenarios where the coupling between the stellar magnetic field and the planetary one lead to an increased drag area
and therefore infall of the planet as compared to the case of an unmagnetized planet. The finding that our observations
most likely imply a planetary magnetopause may turn out to be important here and motivate studies of planetary evolution in
late-type stars that include magnetic star-planet interaction (omitted by Villaver \& Livio, 2007, but included by 
Nordhaus \& Blackman, 2006 who predict stellar outflows). The case for exoplanetary
magnetism proposed here can only be strengthened by a combination of extensive simultaneous polarization monitoring and
repeated spatially resolved observations. The experimentum crucis for an unambiguous evidence of planets will be the
detection of orbital motion in the proper motion of maser spots showing the magnetic fluctuations, by 
contemporaneous direct imaging of the suggested wake flows at the same velocity as the fluctuations. Such a double
evidence would rule out ambiguities: there may be planets without wake flows, if the wind density is not sufficient to
produce observable features, and there are radial alignments of maser spots not related to wake flows. The coincidence of
a persisting pseudo-periodic magnetic fluctuation with a radial alignment of maser spots strongly suggests that these
phenomena are associated and naturally explained by a precessing Jovian magnetosphere and a planetary wake flow,
respectively.
\begin{acknowledgements}
     Based on observations with the IRAM 30m telescope. IRAM is supported by INSU/CNRS (France), MPG (Germany)
     and IGN (Spain). Gabriel Paubert, supported by the IRAM backend group, built VESPA which provides simultaneous
     measurements of all Stokes parameters. We used the valuable optical lightcurves of the AAVSO (http://www.aavso.org)
     and acknowledge support from the telescope staff. We thank the anonymous referee for his constructive and helpful
     comments. We dedicate this paper to the memory of Matthew Carter, who passed away in November 2008. Matt was a valued
     colleague and friend who made many important contributions to the success of the IRAM receivers.
\end{acknowledgements}
\begin{appendix}
\section{Determination of orbital parameters}
Here we summarize the determination of orbital parameters, which follows the classical
method of analysing spatially and spectrally resolved binary systems. Without further
observational evidence at hand, we only treat the case of a circular orbit here.
   \begin{figure}
   \centering
        \resizebox{12cm}{!}{\includegraphics{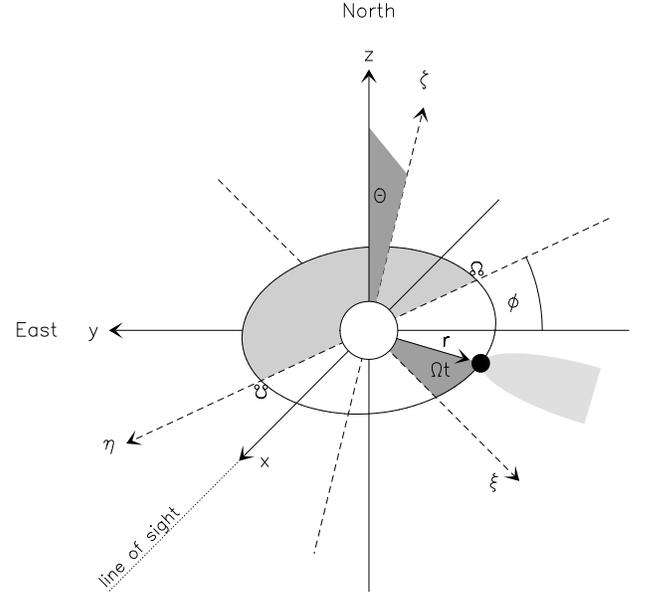} \hspace{4.0cm}}
        \caption{Definition of the orbital parameters. The $xyz$ coordinate system is shown by
	solid arrows (the $yz$ plane, i.e. the drawing plane, is the plane of the sky), the $\xi\eta\zeta$
        coordinate system by the dashed arrows (the orbit is in the $\xi\eta$ plane). The $\eta$ axis is the node
        line about which the orbital plane is inclined by angle $\Theta$ and lies in the $xy$ plane. $\Phi$ is the
        longitude of the ascending node. The star, the planet and its wake flow are schematically indicated, as
        well as the position of the ascending and of the descending node.}
   \end{figure}
The coordinate systems applied and the orbital parameters are defined in Fig.~A.1.
The transformation of the radius vector (pointing from the star to the planet) between the
orbital plane which defines the system $\xi\eta\zeta$ and the $xyz$ coordinate system (where the
$x$ axis is the line of sight to the observer, and where the $yz$ plane is the sky plane) is given by
\begin{equation}
\overrightarrow{r} = \left( \begin{array}{c}
                    X \\
		    Y \\
		    Z 
                 \end{array}    \right) = r. \left( \begin{array}{l}
                                +\cos{\Theta}\cos{\Phi}\cos{(\Omega\Delta t)}-\sin{\Phi}\sin{(\Omega\Delta t)} \\ 
                         	-\cos{\Theta}\sin{\Phi}\cos{(\Omega\Delta t)-\cos{\Phi}\sin{(\Omega\Delta t)}} \\
                         	-\sin{\Theta}\cos{(\Omega\Delta t)}
                       	    \end{array} \right)
\label{eq:r}
\end{equation}
where $r$ is the distance between star and planet, $\Theta$ is the angle between the orbital angular
momentum vector and the direction pointing from the star northwards (taken positive if the positive $\zeta$ axis is
tilted in direction of the positive $\xi$ axis), and $\Omega = 2\pi/T$ is the angular velocity of the planet's orbit
(with period $T$). The angles $\Phi$, $\Theta$ and $\Omega\Delta t$ are the three Euler angles of the transformation
between the $\xi\eta\zeta$ and $xyz$ coordinate systems. They are related to the inclination $i$ of the orbital angular
momentum vector with respect to the line-of-sight to the observer by
\begin{equation}
\cos{i} = \sin{\Theta}\cos{\Phi}
\label{eq:inc}
\end{equation}
$\Delta t = t-t_0$ is the time lapse between the epoch of the observation and the reference time $t_0$
when the radius vector is along the vector pointing from the star to the ascending node
(the node line defines the axis about which the orbital plane is inclined with respect to the line
of sight). The $xyz$ coordinate system is stationary with respect to the local standard
of rest (LSR). Applying the same coordinate transformation to the velocity vector of the planet
$\overrightarrow{v_{\rm P}}$ yields
\begin{equation}
\overrightarrow{v_{\rm P}} = \overrightarrow{v_\ast}+v_{\rm K} \left( \begin{array}{l}
                                -\cos{\Theta}\cos{\Phi}\sin{(\Omega\Delta t)}-\sin{\Phi}\cos{(\Omega\Delta t)} \\ 
                                -\cos{\Theta}\sin{\Phi}\sin{(\Omega\Delta t)}-\cos{\Phi}\cos{(\Omega\Delta t)} \\ 
                         	+\sin{\Theta}\sin{(\Omega\Delta t)}
                       	    \end{array} \right)
\label{eq:v}
\end{equation}
where $\overrightarrow{v_\ast}$ is the proper motion vector of the star with respect to the LSR
(in astronomical velocity definition, i.e. negative velocity corresponds to motion
towards the observer). Here $v_{\rm K}$ is the Keplerian speed of the planet on the
assumed circular orbit, i.e.
\begin{equation}
v_{\rm K} = \Omega .r = \sqrt{\frac{GM_\ast}{r}}\,,
\label{eq:vK}
\end{equation}
where $M_\ast$ is the stellar mass. For the precision required here, it is sufficient to
assume that the center of gravity of the star-planet system is undistinguishably close
to the center of the star.

The analysis of the observations is straightforward now: We need to identify a predominantly
radial alignment of SiO maser spots at the velocity where the fluctuations of circular
polarization are observed, suggesting that they are located in the wake flow of a planet
whose magnetic dipole axis is misaligned with its rotation axis. The velocity at which the
fluctuations occur is interpreted as the line-of-sight velocity of the planet in the
LSR, $(\overrightarrow{v_{\rm P}})_{\rm x}$. A sine wave fit to the measurements of
\begin{eqnarray}
\overrightarrow{(v_{\rm P}})_{\rm x} & = & (\overrightarrow{v_\ast})_{\rm x}+ v_{\rm K}.(\cos{\Theta}\cos{\Phi}\sin{(\Omega\Delta t)-\sin{\Phi}\cos{\Omega\Delta t}} \nonumber \\
                          & = & (\overrightarrow{v_\ast})_{\rm x}+V.\sin{(\Omega\Delta t +\psi)} \hfill
\label{eq:vPlsr}
\end{eqnarray}
at several epochs yields the stellar LSR velocity $(\overrightarrow{v_\ast})_{\rm x}$ and the angular frequency
$\Omega$, as well as the parameters
\begin{eqnarray}
V & = & v_{\rm K}\sqrt{1-\cos^2{\Phi}\sin^2{i}} {\,\,\,\rm and} \hfill \nonumber\\
\tan{\psi} & = & \tan{\Phi}\sec{i}\,. \hfill
\end{eqnarray}
These parameters are used together with the positions of the wake flows in the VLBI maps to 
constrain the radius of the orbit, $r$, and therefore the Keplerian velocity $v_{\rm K} = \Omega r$,
as well as the longitude of the ascending node, $\Phi$, and the inclination of the
orbit with respect to the sky plane, from Eq.\,\ref{eq:r}. The stellar mass is then determined from 
\ref{eq:vK}. As positions, we use the head (i.e. the substellar tip) of the wake flow which we identify
with the planet's position. The corresponding velocity is the maximum speed detected at the head of
the wake flow, since the AGB wind is decelerated downstream the wake flow.

The main uncertainty of this combined kinematical and astrometrical method is the association 
between a magnetic fluctuation and a precessing planetary magnetosphere, and between the
planet's position and the head of the wake flow. It is therefore advisable to over-determine
the velocity and position fit, in order to minimize the observational uncertainties inherent to the
method.
\end{appendix}

\end{document}